# Photostimulation activates restorable fragmentation of single mitochondrion by initiating oxide flashes

**Running title: Photostimulated mitochondrial dynamics**

Yintao Wang[1], Hao He[1,2*], Shaoyang Wang[1], Yaohui Liu[3], Minglie Hu[1], Youjia Cao[3], Chingyue Wang[1*]

[1] Ultrafast Laser Laboratory, Key Laboratory of Optoelectronic Information Technology

(Ministry of Education), College of Precision Instrument and Optoelectronics Engineering,

Tianjin University, Tianjin 300072, P.R. China.

[2] Med-X Research Institute and School of Biomedical Engineering,

Shanghai Jiao Tong University, Shanghai 200030, China

[3] Key Laboratory of microbial functional genomics of Ministry of Education,

College of Life Sciences, Nankai University, Tianjin 300072, P.R. China.

*Corresponding author: haohe@sjtu.edu.cn

*Email: chywang@tju.edu.cn

**Mitochondrial research is important to ageing, apoptosis, and mitochondrial diseases. In previous works, mitochondria are usually stimulated indirectly by proapoptotic drugs to study mitochondrial development, which is in lack of controllability, or spatial and temporal resolution. These chemicals or even gene techniques regulating mitochondrial dynamics may also activate other inter- or intra-cellular processes simultaneously. Here we demonstrate a photostimulation method on single-mitochondrion level by tightly-focused femtosecond laser that can precisely activate restorable fragmentation of mitochondria which soon recover their original tubular structure after tens of seconds. In this process, series of mitochondrial reactive oxygen species (mROS) flashes are observed and found very critical to mitochondrial fragmentation. Meanwhile, transient openings of mitochondrial permeability transition pores (mPTP), suggested by oscillations of mitochondrial membrane potential, contribute to the scavenging of redundant mROS and recovery of fragmented mitochondria. Those results demonstrate photostimulation as an active, precise and controllable method for the study of mitochondrial oxidative and morphological dynamics or related fields.**

**Introduction**

Mitochondrial dynamics have long been deemed as a significant tache in cell life for regulating series of vital processes such as respiration, aging and apoptosis (Ferri & Kroemer, 2011; Sahin & DePinho, 2012; Beal, 2005). Moreover, mitochondrial lesion is recognized to be involved in series of neurodegenerative diseases (Lin & Beal, 2006; Calvo & Mootha, 2010), particularly, Parkinson's disease (Zima & Blatter, 2006; Henchcliffe & Beal, 2008). In previous works, to study the mechanisms of mitochondrial pathology and response to stress (McFarland et al, 2010; Richard & van der Bliek, 2012; Yan et al, 2013), mitochondria are usually stimulated by non-specific chemicals or even proapoptotic drugs (Abou-Sleiman et al, 2006; Tufi et al, 2014). For example, superoxide flashes can be generated in mitochondria in the early stage of stress-induced apoptosis (Ma et al, 2011), and gene deficiency of PINK1 causes mitochondrial accumulation of calcium, excessive reactive oxygen species (ROS) production and opening of mitochondrial permeability transition pores (mPTP), and finally death of neurons with $Ca^{2+}$ treatment (Gandhi et al, 2009). It should be noted that such biochemical methods are in lack of spatial and temporal resolution, and controllability at single-mitochondrion level (Karaman et al, 2008). In this regard, it is challenging to study mitochondrial dynamics with precise and controllable stimulation in a live cell and to bring breakthroughs on researches of mitochondria (Galluzzi et al, 2012).

In recent years, laser technology has provided biological research with precise optical disruption techniques including optical transfection (Tirlapur et al, 2002), cell-cell fusion (He et al, 2008), and neuroaxotomy (Yanik et al, 2004). The ultrahigh peak power and ultrashort pulse width of femtosecond laser make it possible to disrupt targeted organelle without providing significant damage to the whole cell (Vogel et al, 2005). This inspired us of developing a precise and controllable optical method to stimulate mitochondria. Different from laser-induced disruption of cell structures (Shimada et al, 2005), we found that tightly-focused femtosecond laser (Ti:Sapphire mode-lock laser, 810 nm, 75 fs, 80 MHz) could offer controllable photostimulation to initiate restorable fragmentation of single mitochondrion by activating mitochondrial ROS (mROS) flashes and transient mPTP openings with interesting mechanism. We propose that this method can be used for the in-depth research on mitochondrial oxide flash (Wang et al, 2008), mito-fission (Richard & van der Bliek, 2012), or other topics of related areas (such as ageing, oxidative related diseases and neuroscience).

Results

**Photostimulation can induce restorable fragmentation of single mitochondrion**

HeLa cells were used as the proof-of-principle demonstration in this study. In experiments, the laser beam was expanded by a pair of lenses for the collimation and matching to the objective back aperture. This can be also used to control the focus position in the vertical direction by tuning the distance between the two lenses. The laser focus was less than 1.5 μm in diameter to ensure that only one single mitochondrion gets photostimulation at one exposure (Watanabe et al, 2004). Targeted mitochondria in cells were selected randomly and then moved to the position of laser focus by tuning the microscope specimen stage. In experiments, photostimulation was performed at 6 mW (at cell) for 100 ms, the same power level of multiphoton microscopy, which was traditionally thought to have little damage or perturbation to cells (Zipfel et al, 2003). Interestingly, immediately after the laser stimulation, morphology of stimulated mitochondria changed (n=59,

Fig 1A): most fragmented (n=56, Figure E1A) while the rest swelled (n=3). However, the neighboring mitochondria were not influenced (Figure E1A). It took only around 10 s for the stimulated mitochondria to develop significant fragmentation after photostimulation (Fig 1B, n=25), much faster than the fragmenting process induced by biochemical methods (Lin & Beal, 2006; Gomes et al, 2011). More interestingly, the fragmentation of the stimulated mitochondria was restorable. It took only around 40 s for fragmented mitochondria to recover its tubular structure (Fig 1B, n=21), suggesting this photostimulation as a homeostatic perturbation to mitochondria. Swelled mitochondria were not restorable. Therefore we took swelling as more severe damage in our following studies.

To assay the structural non-connectivity of fragmented mitochondria, mitochondrial matrix–targeted photoactivable green fluorescent proteins (mitoPAGFP) was transfected to cells before experiments (Patterson et al, 2002). Half of the targeted mitochondrion was at first scanned by 405-nm laser for PAGFP activation (white box in Fig 1C) while rest part remained dark. If the mitochondrion was fragmented, the activated mitoPAGFP could not diffuse from the activated region to the dark region (as the "Frag" labeled mitochondrion in Fig 1C and Figure E2). Otherwise the bright mitoPAGFP (in white box) could diffuse rapidly to its dark part (as the "Control" labeled mitochondrion in Fig 1C). In this way, photostimulated fragmentation of mitochondria was confirmed. After recovery, the activated mitoPAGFP could then diffuse out.

The recovered mitochondria could suffer more times of photostimulation (n=16, Fig 1D). In this way, a single mitochondrion could be stimulated for multi-times and research on mitochondrial stimulation could be thus more maneuverable (Figure E3). The photostimulated mitochondrion was stimulated for the second time after a short rest, and again could it fragment and recover (Fig 1D). The second fragmentation and recovery durations were very similar to those in the first time (Fig 1E) suggesting stimulated mitochondria could recover to homeostasis after photostimulation.

Since heat could diffuse rapidly to a wide extent resulting in a poor spatial resolution, we suspected that multi-photon excitation might contribute to the major part of photostimulation rather than photo-thermal effect in this method. To test this hypothesis, mitochondrial responses to different photostimulation durations (0.1 s, 0.3 s, and 1 s, at 5 mW) were studied. The result showed only restorable fragmentation for all stimulation durations (Fig 1F), suggesting that photostimulation of mitochondria was independent of heating energy and duration. As a positive control, mitochondria (n=10) were irradiated by a continuous-wave laser (808 nm, 10 mW) with greater thermal effect for 3 s, but all mitochondria showed no response (n=10). In contrast, it was found that the peak power and wavelength of femtosecond laser determined responses of the stimulated mitochondria. As shown in Fig 1F, the stimulated mitochondria showed restorable fragmentation at 6 mW, irrevocable fragmentation at 18 mW (fragmented mitochondria could not recover), and was thoroughly disrupted at 30 mW (all for 0.1 s). To further verify this point, another homemade Yb-doped fiber femtosecond laser (50 MHz, 140-330 fs) at 1040 nm was used to stimulate mitochondria with tunable pulse width and peak power (Zhang et al, 2010). Very different from the results by 800-nm laser (Fig 1G), the fiber laser could not stimulate any mitochondrial response at 6 mW. When the pulse width of the 1040-nm laser was tuned from 330 fs (Fig 1G, 12 mW L) to 140 fs (Fig 1G, 12 mW S), the photostimulation effect was significantly enhanced. It should be noted that at 12 mW 140 fs (peak photon density~ $10^{11}$ W/cm$^2$) or 24 mW 330 fs (peak photon density~ $7.5 \times 10^{10}$ W/cm$^2$), the 1040-nm fiber laser could stimulate comparable but still less prominent mitochondrial responses than 810 nm fs-laser at 6 mW (peak photon density~ $5 \times 10^{10}$ W/cm$^2$). These results which strongly depended on peak power (photon density) and wavelength suggested that multiphoton excitation contributed as the major part to photostimulating process. When the Ti:Sapphire laser power was increased to 60 mW, the plasma effect of multiphoton excitation would be very significant to damage the cell (Figure E4).

**Photostimulated mROS oxide flashes and MMP oscillations are critical to mitochondrial restorable fragmentation**

In experiments, it was observed that mROS was generated immediately after photostimulation: at first merely in the laser focal spot, soon in the whole mitochondrion, and finally scavenged (Fig 2A and Figure E5). This burst of mROS resembled the "superoxide flash" which was thought as spontaneous mitochondrial behavior in previous report (Wang et al, 2008). But in our work, mROS flash could be only excited actively by photostimulation. Considering the damage effect of ROS, we suspected that the laser-activated mROS "flash" played a critical role in the mitochondrial morphological response to photostimulation. To test this hypothesis, antioxidant N-Acetyl Cystein (NAC, 5 mM) was used to scavenge (or attenuate) mROS (Carrière, 2009; Roy et al, 2008). It was found that NAC could effectively prevent mitochondrial fragmentation from photostimulation (indicated by significantly lower percentage of fragmented and swelled mitochondria, Fig 2B). As control, if cells were treated with high oxidative stress (tert-Butyl hydroperoxide, i.e. TBHP, 20 μM), the effectiveness of photostimulation was then significantly enhanced (Fig 2B).

Once mROS flash was initiated, the stimulated mitochondrion might generate mROS flashes repeatedly, to which a series of mitochondrial membrane potential (MMP) oscillations corresponded exactly (Fig 2C, n=20 independent experiments, and 4 selected most significant flash events in Fig 2C were shown in Figure E6). Such fast MMP oscillations probably suggested the transient openings of mPTP, which were related with mROS flashes and might contribute to scavenging of mROS and mitochondrial recovery.

We then studied photostimulated MMP oscillations in different redox environments. The ratio of mitochondria with multi-time MMP oscillations (statistics in 100 s after photostimulation) decreased (Fig 2D) in high oxidative-stress environment (TBHP, 20 μM). If the cellular or mitochondrial ROS scavenging ability was enhanced (by anti-oxidants NAC or mitochondria-targeted antioxidant Mito-TEMPO, 5 mM and 5 μM respectively), the ratio of mitochondria with multi-time MMP oscillations then increased (Fig 2D). The reason might be that, with high oxidative stress, the stimulated mitochondria suffered more damage by ROS to induce long-time mPTP opening and probably depolarization of MMP (similar to the initial processes of apoptosis); with anti-oxidant there would be less mROS generation and mitochondria damage, the photostimulation was then more homeostatic with less damage, and the stimulated mitochondria could easily activate fast transient mPTP openings to recover and protect MMP from depolarization. To further support this point, we measured the MMP level 100 s after photostimulation. It could be found that the MMP recovery ratio of mitochondria with anti-oxidant was much higher than it with oxidative stress (Fig 2E).

**mPTP opening plays a key role in mitochondrial recovery**

We further investigated the role of mPTP in the recovery process of photostimulated mitochondria. At first, mPTP opening was assayed by calcein whose fluorescence would be quenched by $Co^{2+}$ diffusing from cytosol into mitochondria if mPTP were permanently open. As shown in Fig 3A, during the restorable fragmentation stimulated at 6 mW, there was no fluorescence quenching indicating no permanent (or long-time) mPTP opening. As control, calcein fluorescence quenching could be observed after intense laser stimulation at 20 mW which induced permanent (or long-time) mPTP opening and unrecoverable fragmentation (or swelling). When mPTP was inhibited by cyclosporin A (CsA, 10 μM), in Fig 3B, it could be found that photostimulation (at 3 or 6 mW) with CsA could induce much higher percentage of fragmented and swelled mitochondria than in the control group (without CsA). More specifically, with CsA, the fragmentation and recovery of photostimulated mitochondria (Fig 3C and Figure E7) were significantly slower than in normal condition (Fig 1B). Therefore, in this study, transient mPTP openings contributed to mitochondrial recovery. It should be noted that there was no ROS or $Ca^{2+}$ activation in cytosol. In this study, mROS were directly generated inside the stimulated mitochondria, which was very different from previous studies where the mROS were stimulated by oxidative or $Ca^{2+}$ stress in cytosol and extracellular stress in buffer. Hence it could be reasoned the transient mPTP openings here then played a role that contribute to scavenging mROS inside the stimulated mitochondrion for mitochondrial recovery. This was different from previous studies, where the mPTP opening would induce oxidative or $Ca^{2+}$ stress invading from cytosol into mitochondria and then mitochondrial stimulation (Ma et al, 2011; Wang et al, 2008). In this regard, there should be an accumulation of mROS in the stimulated mitochondrion after mPTP inhibition even though the stimulation was ultra-weak. To assay this process, mitochondria were then stimulated at only 0.5 mW for a long duration with the presence of CsA. It can be found in Fig 3D that mROS could be slowly accumulated to the same level as the one treated at 3 mW for 0.2 s. Therefore, this ultra-weak (0.5 mW) photostimulation could also induce mitochondrial restorable fragmentation when mPTP was inhibited (Fig 3E). But without CsA, such ultra-weak (long-duration) photostimulation could not activate any mitochondrial response even if increasing the power to 1 mW and stimulation duration to more than 60 s, probably because mPTP could now work for mROS release or scavenging. Hence our results suggested that transient mPTP opening contributed to the protection of mitochondria from photostimulation for their recovery by scavenging mROS.

**Molecular translocation from or to stimulated mitochondria**

We further showed that mPTP were the key translocation regulator of proapoptotic protein Bax or cytochrome c (Cyto C) in the stimulated mitochondria. As shown in Fig 4A and B, when the laser power was at 6 mW, no translocation of Bax or Cyto C was found. But if the power was increased to around 20 mW, concentration of Bax to the stimulated mitochondria and Cyto C release from them could be observed. Thus the photostimulation in this study (6 mW) would not induce any destructive effect to the mitochondria. When the mPTP were permanently opened by intense stimulation at high laser power (20 mW), the molecular translocation only happened locally around those stimulated mitochondria. And their morphological change would not recover (Fig 1F). After 6 hours, all stimulated cells with translocation of those proapoptotic molecules did not go to apoptosis. Hence the localized proapoptotic signals would not influence the cell fate. According to the results above, we summarized photostimulation process as the proposed mechanism in Fig 4C.

**Discussion**

Our experimental results suggested photostimulated mROS were essential to mitochondrial fragmentation and mPTP played a key role in mitochondrial recovery. Different from previously observed spontaneous superoxide flashes (Wang et al, 2008), mROS flashes in our study were stimulated directly and actively, which, however, might provide information to the unclear mechanism of the spontaneous superoxide flashes. The process of how laser stimulation generated mROS remained unknown. Some mitochondrial proteins, complex I-IV for instance, may be damaged or perturbed during the multiphoton excitation of photostimulation, resulting in a fast mROS generation (Liu et al, 2002; St-Pierre et al, 2002). The damaged mitochondria could not maintain normal respiration processes and hence induce series of mROS flashes. Nevertheless, the photostimulation was somehow homeostatic, and the stimulated mitochondria could still repair themselves to recover normal by scavenging mROS. In this process, the transient mPTP openings played a different role with previous studies as the mechanisms of stimulation to mitochondria were different. The precise photostimulation brought little perturbation to the surroundings of the targeted mitochondrion (Figure E8), and ROS was merely generated inside it. Therefore, transient mPTP openings then worked for mROS scavenging, whereas in previous studies ROS and $Ca^{2+}$ in cytosol were the main damaging stress to mitochondria and mPTP openings would induce influx of ROS and $Ca^{2+}$ into mitochondria.

In conclusion, we show here that single mitochondrion can be precisely stimulated by tightly-focused femtosecond-laser irradiation through multiphoton excitation. Most stimulated mitochondria develop to restorable fragmentation and few to swelling. Fragmented mitochondria can recover to homeostasis in a short duration, and can endure multi-time photostimulation after their recovery. The photostimulated mitochondrion can generate series of mROS flashes, which play an important role in mitochondrial morphological change. Transient mPTP openings may contribute to the scavenging process of mROS and thus are very essential to mitochondrial repair (recovery) process. If the photostimulation were too intense, the mitochondria will be greatly damaged and mPTP will be permanently opened to induce localized translocation of proapoptotic proteins such as Bax and CytoC. Based on all these results, we propose that this noninvasive, controllable and precise optical method may serve as a promising stimulation technique for the research of mitochondria in the study of cell biology and mitochondrial pathology in neuroscience or even ageing, oxidative stress-related diseases.

## Materials and Methods

**Optical setup and photostimulation**

The whole system was set on a confocal microscopy (Olympus FV1000/IX81) coupled with a homemade Ti:Sapphire femtosecond laser and a Yb-doped fiber laser. The Ti:Sapphire femtosecond laser was pumped by a green laser diode Millennia V (Spectra-Physics) and had an output power at 400 mW. Both femtosecond laser beams were at first collimated and expended to match the back aperture of objective (~8 mm), then reflected by a beam splitter (reflection=70%) and focused by the objective (water-immersed, 60X, near-infrared coating, N.A.=1.2). A CCD was put under the beam splitter for monitoring the laser focus. In this way, a diffraction-limit laser focus could be achieved. The laser power was controlled by an attenuator and stimulation duration by a mechanical shutter (resolution: 10 ms). The laser focus was hence fixed in the center of field of vision. In the vertical direction, the laser focus was adjusted at the same plane of confocal scanning by tuning the distance of the two lenses for beam collimation to adjust the divergence angle of the collimated laser beam slightly, which could then be focused to different vertical position. This setup was different from two-photon microscope, where the femtosecond laser was also controlled by the galvanometer sharing the same light path with the CW excitation lasers. In our setup, the femtosecond laser beam was not scanned and thus fixed to a single point to get a high spatial resolution.

In experiments, the laser was at first blocked by the mechanical shutter. Once the targeted mitochondrion was selected, it was then moved to the position of laser focus by adjusting the stage of the microscope. The femtosecond lasers and fluorescence and excitation lasers were split by a dichroic mirror (reflection<750 nm; transmission> 750 nm). Therefore, during photostimulation, confocal scanning could be still performed and all information could be recorded.

**Fluorescent confocal microscopy**

TMRM, DihydroRh123, and Mitochondrial Transition Pore Assay Kit (Calcein and $CoCl_2$) were used for MMP, mROS, and mPTP labeling. The staining procedures were followed by the protocols provided by the supplier. When using the mPTP kit, cells were at first loaded with calcein AM and $CoCl_2$. Then after incubation, cells were washed three times and $CoCl_2$ was added to cell buffer to maintain high $Co^{2+}$ concentration in cytosol. The fluorescence from cytosolic calcein was immediately quenched by $CoCl_2$, while the fluorescence from the mitochondrial calcein was maintained (since $CoCl_2$ could not diffuse into mitochondria). But once mPTP open, $CoCl_2$ could then pass through mPTP into mitochondria and quench the fluorescence of calcein there. Hence the decline of calcein fluorescence value inside mitochondria indicates the opening of mPTP.

For confocal scanning, all green dyes were excited by a 473-nm laser and the red ones by a 543-nm laser. To minimize photobleaching, the power of 473-nm laser was working at around 0.8 mW, and the 543-nm laser at 0.1 mW at output (laser powers at specimen can be estimated by multiplying the coupling efficiency at around 20% of the system). Mostly, photos were taken at 512×512 pixels and each frame took around 1.1 s to scan. The fluorescence of NADH (emission: 420-450 nm) was excited directly by a laser at 351 nm as the autofluorescence of NADH.

**Acknowledgements**

We thank Dr. Heping Cheng (Peking University), Dr. Siukai Kong (the Chinese University of HongKong), Dr. A. Nakagawa (Tohoku University), Dr. Ratnesh Lal (University of California San Diego), and Dr. Sailing He (Zhejiang University) for effective discussions and their valuable suggestions. This work was supported by grants from National Natural Science Foundation of China (NSFC) 61108080 to HH and 81171556 to YC. YW and HH contributed equally to this work.

**Conflict of interest**

The authors declare no conflict of interest.

**Figure Legends**

**Figure 1 - Precise and controllable photostimulation to targeted mitochondria.**

    A   Mitochondria were selected randomly and stimulated by femtosecond laser at the 5th second at 6 mW for 0.1 s. Soon most of the stimulated mitochondria showed fragmentation (the first panel) and then recovered. Others swelled (the second panel) without recovery. Surrounding mitochondria were rarely influenced and all cells showed good integrity and viability after 6 hours. Inserts: magnified figure of stimulated mitochondria. Orange arrow: the position of laser focus.

    B   Time statistics of 25 photostimulated mitochondria developing from restorable fragmentation to recovery. Twenty-one fragmented mitochondria recovered within 100 s after photostimulation while it took around 100 s, 200 s and 300 s respectively for the other four.

    C   Fragmentation of stimulated mitochondria was verified by mitochondrial matrix-targeted PAGFP. White box: the photo-activated region scanned by 405-nm laser. Orange arrow: photostimulation. White arrows: "Frag" indicated the photostimulated mitochondrion which PAGFP could not diffuse along and "Control" indicated the mitochondrion without photostimulation in which PAGFP could diffuse. Inserts: magnified figure of the control mitochondrion.

    D   Duo-time stimulation to a single mitochondrion (orange arrow). After the first irradiation at 2.2 s, it can be found that the mitochondrion fragmented at 4.4 s and recovered at 24.4 s. After a short rest, the same mitochondrion was then irradiated again at 64 s and the maximum fragmentation was observed at 117 s and then recovered at 166 s.

    E   Time statistics of mitochondrial fragmentation and recovery duration after first and second photostimulation.

    F   Long-time photostimulation only induced fragmentation without severe damage (first panel, n=5 mitochondria for each duration). But high-power photostimulation significantly damaged or even disrupted mitochondria (second panel, n=11 mitochondria for each power. Orange arrow: the position of laser focus.

    G   Photostimulation to mitochondria depended on laser wavelength and peak power. At 1040 nm, the fiber laser could initiate no mitochondrial response at 6 mW even though the thermal effect was 20-time higher than at 810 nm. At 12 mW, the

fragmented and swelled mitochondria stimulated at 140 fs (12 mW S, n=11) took significantly higher percentage than at 330 fs (12 mW L, n=11). At 24 mW (330 fs, n=11), proportions of fragmented and swelled mitochondria were comparable to (but still less than) those at 810 nm, 6 mW. Bar: 10 μm.

**Figure 2 - Laser-induced mROS flashes played a key role in mitochondrial response.**

A  Photostimulated mROS was generated around the laser focus (orange arrow, 8.8 s), and soon spread to the whole mitochondrion (12.1 s), then finally scavenged (42 s).

B  Responses of the photostimulated mitochondria were sensitive to oxidative stress. Mitochondrial restorable fragmentation induced by photostimulation (orange arrow) was greatly inhibited by NAC (5 mM) scavenging ROS (n=11). In contrast, more mitochondria showed swelling and fragmentation in the presence of TBHP (n=25). As control, most mitochondria showed restorable fragmentation after photostimulation (n=59).

C  Photostimulation induced MMP oscillations (labeled by TMRM, red) and mROS flashes (labeled by DihydroRhodamine123, green). Box: the stimulated mitochondrion. Right: MMP oscillation and mROS flashes in the stimulated mitochondrion (Box). Each oscillation of MMP corresponds exactly with a flash of mROS, which probably indicates a transient mPTP opening. Orange arrow: photostimulation. Arrows with numbers: selected four significant events of mROS flashes.

D  The ratio of photostimulated mitochondria with different MMP oscillation times (in 100 s) was related with mitochondrial oxidative environment. More mitochondria showed multi-time transient MMP oscillations under the presence of NAC (5 mM) or Mito-TEMPO (5 μM) while fewer MMP oscillations could be found in mitochondria treated with TBHP (20 μM). (e) The MMP recovery ratio with anti-oxidants was much higher than the TBHP-treated group 100 s after photostimulation. Depolarized: depolarized mitochondria. Bar: 10 μm.

**Figure 3 - mPTP played a key role in mitochondrial recovery.**

A  Different types of mPTP opening varied with different photostimulation power. The opening of mPTP was monitored by calcein, whose fluorescence would be quenched by $Co^{2+}$ that could not diffuse into mitochondria unless mPTP were long-time opened. No permanent (long-time) mPTP opening could be observed when the laser power was at 6 mW. At 20 mW, the photostimulation could damage mitochondria significantly to induce permanent mPTP opening and the calcein fluorescence was then quenched. Left: mitochondria before and after photostimulation. Right: calcein fluorescence of the stimulated mitochondria. Orange arrow: the position of laser focus. Bar: 10 μm.

B   Higher percentage of fragmented or swelled mitochondria were stimulated with the presence of CsA.

C   If mPTPs were inhibited by CsA, duration for mitochondrial recovery was significantly elongated.

D   With the presence of CsA, ultra-weak photostimulated (0.5 mW for 30 s) mitochondrion witnessed a slow mROS accumulation which was comparable to the level stimulated at 3 mW for 0.2 s (without CsA). Without CsA, no mROS increase could be found at 0.5 mW for any duration.

E   Restorable fragmentation could be initiated by such ultra-weak stimulation at 0.5 mW for 30 s with the presence of CsA (n=15). Without CsA, no response could be found even at 1 mW. Orange arrow: the position of laser focus. Bar: 10 μm.

**Figure 4 - Molecular translocation in photostimulated mitochondria.**

A   Bax would not translocate at 6 mW (left, n=5) but would concentrate to the stimulated mitochondria at 20 mW (right, n=5). IF: immunofluorescence microscopy, which was performed 20 minutes after photostimulation (orange arrow). MMP would not be depolarized significantly at 6-mW stimulation. Orange arrow: the position of laser focus. Bar: 10 μm.

B   Similarly, release of Cyto C from the stimulated mitochondrion would only take place after 20-mW photostimulation (orange arrow, n=5). Orange arrow: the position of laser focus. Bar: 10 μm.

C   Proposed mechanism of mitochondrial response to photostimulation. At 6 mW, photostimulated mROS generation induced restorable fragmentation and transient mPTP opening, which could release mROS for recovery and might repeat several times automatically resulting in MMP oscillations and mROS flashes. At 20 mW, such intense photostimulation generated great damage to the mitochondrion and thus induced permanent opening of mPTP, which activated release of Cyto C from the targeted mitochondrion and translocation of Bax to it. Dashed arrows: diffusion of molecules.